\documentclass[prb,twocolumn,showpacs,preprintnumbers,amsmath,amssymb]{revtex4}

\usepackage{graphicx}
\usepackage{dcolumn}

\newcommand{\beq}{\begin{equation}}
\newcommand{\eeq}{\end{equation}}
\newcommand{\beqa}{\begin{eqnarray}}
\newcommand{\eeqa}{\end{eqnarray}}

\newcommand{\qvec}{{\bf q}}

\begin{document}
\title{Disorder effects in the quantum Heisenberg model:
An Extended Dynamical mean-field theory analysis}

\author{S. Burdin$^{1}$, D. R. Grempel$^{2}*$, and M. Grilli$^{3}$}

\affiliation{$^1$  Max-Planck-Institut f\"ur Physik komplexer Systeme. 
N\"othnitzer Stra$\beta$e 38, 01187 Dresden, Germany}
\affiliation{$^2$ CEA-Saclay$/$DSM$/$DRECAM$/$SPCSI, 91191 Gif-sur-Yvette Cedex,
France}
\affiliation{$^3$INFM-CNR
 SMC Center, and Dipartimento di Fisica\\
Universit\`a di Roma "La Sapienza" piazzale Aldo Moro 5, I-00185 Roma, Italy}

\begin{abstract}
We investigate a quantum Heisenberg model with both antiferromagnetic
and disordered nearest-neighbor couplings. We use an extended dynamical
mean-field approach, which reduces the lattice problem to a self-consistent
local impurity problem that we solve by using a quantum Monte Carlo algorithm.
We consider both
two- and three-dimensional antiferromagnetic spin fluctuations and systematically
analyze the effect of disorder. We find that in three dimensions
for any small amount of disorder a spin-glass phase is realized. 
In two dimensions, while clean systems display the properties of a highly 
correlated spin-liquid (where the local spin susceptibility
has a non-integer power-low frequency and/or temperature dependence),
in the present case this behavior is more elusive
unless disorder is very small. This is because the spin-glass 
transition temperature leaves only an intermediate temperature regime where
the system can display the spin-liquid behavior, which turns out to be more apparent
in the static than in the dynamical susceptibility.

\end{abstract}
\pacs{75.10.Jm, 75.10.Nr, 75.50.Ee, 75.40.Gb}
\date{\today}
\maketitle

\section{Introduction}

Quantum magnetism is an important topic of modern solid state
physics: Not only do magnetic phases take place in the phase diagram of
many correlated electron systems, like high-temperature superconducting
cuprates \cite{reviewcuprates} and heavy fermions \cite{reviewHF}, 
but many different materials display
magnetic properties, which change upon varying parameters such as 
doping, temperature, disorder, or pressure. A few examples 
may (very partially) illustrate the enormous variety 
of behaviors that one encounters in this field: Upon increasing doping
the lightly doped ${\rm La_{2-x}Sr_xCuO_4}$ passes from an antiferromagnetic
 (AF) to a spin-glass (SG) phase\cite{reviewcuprates}; the introduction of non-magnetic
impurities in quantum paramagnets like ${\rm SrCu_2O_3}$, ${\rm KCuCl_3}$,
 ${\rm CuGeO_3}$, gives rise to local-moment formation\cite{webervojta}; the glassy phase
of the disordered dipolar-coupled magnet ${\rm LiHo_xY_{1-x}F_4}$ may
be driven to other magnetic phases by means of a magnetic field\cite{LHYF};
${\rm SrCr_{9p}Ga_{12-9p}O_{19}} $ is a $S=3/2$  magnet with a Kagom\'e lattice structure,
which above the (low) SG freezing temperature displays the spin-liquid behavior 
\cite{notespin-liquid,sachdevye} typical of strongly correlated spin systems\cite{mondelli}. 
From this large variety we extract three essential ingredients, which
are commonly and physically relevant in quantum magnetic materials: disorder, frustration, and
fluctuations (both thermal and quantum). The separate role of these physical mechanisms 
and their interplay in determining different physical properties is 
obviously a relevant issue in the field of quantum magnetism.
Since one is facing a huge variety of systems with different structures, properties
and phase diagrams, the search of common fundamental mechanisms mostly relies on the
analysis of schematic models to highlight the deep physical effects. For instance
some work has been devoted to the interplay between
disorder and fluctuations \cite{georgesparcollet,rozenberg,daniel-marcelo}, 
analyzing the role of quantum spin fluctuations in driving a 
spin-glass (SG) phase into a paramagnetic (or spin-liquid)
 phase upon varying spin size and/or temperature.
Within the same framework we consider in this paper a ``simple'' quantum 
AF Heisenberg model  for $S=1/2$ spins and we introduce a random magnetic coupling.
By solving this Disordered Quantum
Antiferromagnetic Heisenberg (DQAFH) model within the Extended
Dynamical Mean-Field Theory (EDMFT)\cite{smithsi,chitrakotliar,BGG}, 
we aim to clarify the interplay 
between AF and random magnetic couplings as well as the role of 
dimensionality in the instability from a paramagnetic to a SG
phase and the relevance of quantum fluctuations in suppressing
the critical temperature. Throughout the present work we will always
remain in the paramagnetic phase with neither AF- nor replica-symmetry breaking
and we will determine the conditions for these
(second-order) instabilities to take place. 

The EDMFT technique \cite{smithsi,chitrakotliar,BGG} is an extension of the
standard Dynamical Mean-Field Theory \cite{reviewDMFT}, where the local 
(spin or charge) degrees of freedom are coupled to a self-consistent bosonic
bath. In our case the Heisenberg model is mapped onto a single
quantum spin embedded in a bosonic bath which 
represents the surrounding spin excitations as a fluctuating magnetic field.
This approach has some definite limitations: 
It is not suited to keep full account of the spin character of
the surrounding degrees of freedom (for instance it misses the possibility
of singlet formation between the single spin and the surrounding ones)
and it neglects the momentum dependence of the spin self-energy thereby 
preventing the occurrence of spatial anomalous dimensions. This
latter limitation is more severe at low temperatures, where spatial
correlations between the spins become relevant. As also discussed in Section VI 
the consequence of this 
is the appearance of well-known \cite{BGG,haule} low-temperature instabilities 
occurring as an unphysical sign in the spin self-energy. Also related to this
poor treatment of spatial correlations (and the related lack of anomalous dimensions)
is the appearance of spurious first-order transitions, where second-order 
are instead expected \cite{PKM}. Nevertheless, it is commonly recognized 
\cite{PKM} that the EDMFT approach is a valuable tool to investigate
the dynamics of spin systems and it makes
an important first step toward the inclusion of spatial correlations
between the spins. 

Our analysis of the DQAFH not only has a broad validity and 
addresses general theoretical issues, but it
is also of pertinence for those numerous AF materials, where disorder
can induce competition between antiferromagnetism (possibly having a correlated
anomalous character) and a spin glass (SG) phase, or can induce local moment formation
with low-temperature Curie-like dependencies.
The role of low dimensionality in determining the relevance of magnetic fluctuations
 is also investigated.

The paper is organized as follows. In Section II we present the model and the
EDMFT technique. Section III presents the criteria necessary to detect the establishment
of broken-symmetry phases or the spin-liquid behavior. The numerical
results for static properties in the three- and 
two-dimensional case are reported in Section IV, while the dynamical 
behavior is discussed in Sect. V.
In Section VI we discuss our results and present our conclusions.

\section{The model and the technique}
\subsection{The model}

We consider the Disordered Quantum Antiferromagnetic Heisenberg (DQAFH) model, 
given by the following Hamiltonian: 
\begin{equation}
\label{hamiltonian}
H=\sum_{i<j}\tilde{J}_{ij}{\bf S}_{i}{\bf S}_{j}~. 
\end{equation}
Here ${\bf S}_{i}$ is a spin$-1/2$ operator at the $i-$th site of a 
lattice of coordination $z$. 
The nearest-neighbor magnetic couplings 
$\tilde{J}_{ij}\equiv \tilde{J}_{ij}^{AF} + \tilde{J}_{ij}^{D}$
are the sum of a translationally invariant part 
$\tilde{J}_{ij}^{AF}\equiv {J}_{ij}^{AF}/\sqrt{z} \equiv J/\sqrt{z}$ [with 
Fourier transforms $J({\bf q})$] and a 
disordered part $\tilde{J}_{ij}^{D}\equiv J_{ij}^{D}/\sqrt{z}$. 
The translationally invariant couplings are related to the spin-wave 
spectral density 
$\rho (\epsilon)\equiv \sum_{{\bf q}}\delta(\epsilon-J({\bf q}))$. 
The disordered parts $J_{ij}^{D}$ are random couplings with 
a Gaussian probability distribution $P(J_{ij}^{D})$, with zero mean value
$\langle J_{ij}^{D} \rangle_{dis}=0$ and 
$\langle J_{ij}^{D} J_{kl}^{D}  \rangle_{dis}=\delta_{i,l}\delta_{j,k}J_D^2$
[$\langle \cdots \rangle_{dis}$ represents the averaging over the 
disorder distribution]. 

In the following, we will perform a large-$z$ expansion, in order to 
study the spin-wave correlations within a single-site effective 
model. This derivation has been described in Ref. \onlinecite{smithsi}
for a clean Heisenberg model. Here, the presence of disorder requires a 
specific preliminary treatment. 

\subsubsection{Disorder and replicas}

The free energy is averaged with respect to the probability 
$P(J_{ij}^{D})$. We use the well-known replica trick\cite{Replicatrick}, 
which relies 
on the fact that the free energy ${\cal F}$ can be calculated from
the partition function ${\cal Z}$ using the relation 
\begin{equation}
\label{Freeenergy}
\beta{\cal F}=\langle \log {\cal Z}\rangle_{dis}
=
\lim_{n\rightarrow 0^{+}}\frac{1}{n}\log \langle {\cal Z}^n\rangle_{dis}~. 
\end{equation}
In order to calculate $\langle {\cal Z}^n\rangle_{dis}$ we introduce 
$n$ replicas of the system, each one being represented by an index $\alpha$ 
\begin{equation}
\label{zndis}
\langle {\cal Z}^{n}\rangle_{dis}
=\prod_{i<j}\int_{-\infty}^{+\infty}dJ_{ij}^{D}P(J_{ij}^{D})
Tr\left[ \exp -\beta\sum_{\alpha=1}^{n}H_\alpha\right]~, 
\end{equation}
where $H_\alpha$ is the disordered Heisenberg Hamiltonian Eq. (\ref{hamiltonian})
acting on the replica $\alpha$. 
In Eq.(\ref{zndis}), the trace is applied to all the operators 
${\bf S}_{i}^{\alpha}$ of the replicated Hilbert space (see Appendix A).

\subsubsection{Extended Dynamical Mean Field}

In the standard DMFT technique\cite{reviewDMFT}, 
the electronic degrees of freedom of a specific single site
are coupled to a fermionic effective bath. On the other hand, in the EDMFT
spin (or charge) degrees of freedom are coupled to an external bosonic
bath simulating the spin/charge degrees of freedom of the rest of 
the lattice \cite{smithsi,chitrakotliar,BGG}. 
The condition that the single specific site must be equivalent
to all other sites of the lattice is then implemented by means of
selfconsistency equations.
We are now going to suitably adapt this procedure to the disordered spin
system represented in Eq. (\ref{hamiltonian}).
The first step is standard: We single out a specific 
site of the system (arbitrarily chosen as the origin); the disorder 
($J_{ij}^{D}$ couplings) will be formally averaged and the trace will be taken 
with respect to the operators ${\bf S}_{i}^{\alpha }$ with $i\neq 0$. 
For any value of $z$, one obtains within this procedure an effective
local action for the local spins ${\bf S}_{0}^{\alpha }$. 
The local action is calculated with a cumulant expansion\cite{burdin}: In the
large-$z$ limit, all the cumulants vanish, except  the
first two terms. 
One depends linearly on ${\bf S}_{i}^{\alpha }$ and corresponds 
to the static mean field. The second is quadratic and reflects the 
dynamical fluctuations. Here, we study only the paramagnetic and 
replica-symmetric solution with $\langle {\bf S}_{i}^{\alpha }\rangle=0$:
The transition to an AF phase or to a SG phase will 
only be analyzed from the paramagnetic phase, considering the criteria 
for a second-order transition  (see Sect. III). 

The local effective action is thus cast into the following form: 
\begin{equation}
\label{complicatedaction}
{\cal A}
=
-\frac{1}{2}
\int_{0}^{\beta}d\tau\int_{0}^{\beta}d\tau '
\sum_{\alpha \alpha '}
{\bf S}_{0}^{\alpha}(\tau){\bf S}_{0}^{\alpha '}(\tau ')
{\cal K}^{\alpha \alpha '}(\tau-\tau ')~, 
\end{equation}
with 
\begin{equation}
{\cal K}^{\alpha \alpha '}(\tau-\tau ')
=\frac{1}{3z}\sum_{ij}\langle J_{0i}J_{j0}
\langle{\bf S}_{i}^{\alpha}(\tau)
{\bf S}_{j}^{\alpha '}(\tau ')
\rangle^{cavity}\rangle_{dis}~. 
\label{KAAbis}
\end{equation}
In this expression for the Kernel, as in the following ones, the summation with
respect to $i$ and $j$ concerns the nearest neighbors of the site $0$, while
$\alpha,\alpha'$ are replica indexes.
The cavity correlation functions $\langle \cdots\rangle^{cavity}$ 
correspond to the correlation functions calculated with a cavity 
instead of the site $0$ (see Appendices A and B). 
The coupling factors $J_{0i}J_{j0}$ in Eq.~(\ref{KAAbis}) precisely 
invoke the site $0$ (of the cavity). 

Using the arguments of Appendix B, the averaging over the disorder 
can be factorized 
\begin{eqnarray}
&&\langle J_{0i}J_{j0}
\langle{\bf S}_{i}^{\alpha}(\tau)
{\bf S}_{j}^{\alpha '}(\tau ')
\rangle^{cavity}\rangle_{dis}
= \nonumber\\
&&\langle J_{0i}J_{j0}\rangle_{dis}
\langle\langle{\bf S}_{i}^{\alpha}(\tau)
{\bf S}_{j}^{\alpha '}(\tau ')
\rangle^{cavity}\rangle_{dis}~. 
\label{PartofKAA}
\end{eqnarray}

Since we are concerned with (the instabilities of) the paramagnetic phase 
we only consider averages (with and without cavities) leading
to diagonal quantities both in the spin components and in the
replica indexes. All the following arguments could be carried out
keeping the matrix character of the spin correlation functions, but
we adopt this simplified presentation to put more clearly in evidence
the role of disorder without unnecessary formal complications.
In this way all thermally averaged quantities are 
scalar quantities and accordingly
\begin{equation}
{\cal K}(\tau-\tau ') \equiv
{\cal K}^{\alpha \alpha}(\tau-\tau ')~,
\label{KAAscalar}
\end{equation}
and one can introduce the (diagonal, i.e., scalar) spin susceptibility
\begin{equation}
\chi_{ij}(\tau-\tau ')=
\langle {\bf S}_{i}^{\alpha}(\tau){\bf S}_{j}^{\alpha}(\tau ') \rangle/3=
\langle S_{i}^{\alpha \mu}(\tau)S_{j}^{\alpha \mu}(\tau ') \rangle~.
\end{equation}
[$\mu$ is a spin-component index].
We also define the disorder averaged susceptibilities 
\begin{equation}
\overline{\chi}_{ij}\equiv \langle {\chi}_{ij}\rangle_{dis}~, 
\label{defchiaverage}
\end{equation}
with a similar definition for the average cavity susceptibilities 
$\overline{\chi}_{ij}^{cavity}$. 
Now, invoking Eq.~(\ref{PartofKAA}) in Eq.~(\ref{KAAbis}), 
fixing the unnecessary replica index and dropping it, we find 
for the paramagnetic phase 
\begin{equation}
{\cal K}(\tau)
=
\frac{1}{z}\sum_{ij}\langle J_{0i}J_{j0}\rangle_{dis}
\overline{\chi}_{ij}^{cavity}(\tau)~. 
\label{Kernelintermediaire}
\end{equation}
In the large$-z$ limit, 
the averaged cavity susceptibility can be expressed in terms of the full 
physical (i.e. without cavity) averaged susceptibility (see Appendix B)
\begin{equation}
\overline{\chi}_{ij}^{cavity}(\omega_{n})
=
\overline{\chi}_{ij}(\omega_{n}) - 
\overline{\chi}_{i0}(\omega_{n})\overline{\chi}_{0j}(\omega_{n})
/\overline{\chi}_{00}(\omega_{n})~. 
\label{averagechicavity}
\end{equation}
where $\omega_n$ is a Matsubara frequency.
The Kernel (\ref{Kernelintermediaire}) is then given by the self-consistent 
relation 
\begin{equation}
\label{completekernel2}
{\cal K}
=
\frac{1}{z}\sum_{ij}
\left( {J_{0i}^{AF}J_{j0}^{AF}+J_{D}^{2}\delta_{ij}}\right)
\left[ {
{{\overline\chi}}_{ij}
-
{{\overline\chi}}_{i0}
{{\overline\chi}}_{0j}/
{{\overline\chi}}_{00}
}\right]~, 
\end{equation}
where the explicit Matsubara frequency dependency has been dropped for
clarity. Using standard considerations \cite{nota1suz,reviewDMFT,smithsi} of 
power counting estimates of the correlation function dependence
on $1/z$ one can cast the part proportional to $J_D^2\delta_{ij}$
in a fully local form
\begin{eqnarray}
\label{completekernel}
{\cal K}
&=&
\frac{1}{z}\sum_{ij}
 J_{0i}^{AF}J_{j0}^{AF}\left(
\overline\chi_{ij}- \overline\chi_{i0}\overline\chi_{0j}/\overline\chi_{00}
\right)
+\frac{1}{z}\sum_{i}J_{D}^{2}\overline\chi_{ii}
\nonumber \\
&\equiv& 
{\cal K}_{AF}+{\cal K}_D~. 
\label{finalkernel}
\end{eqnarray}
Although the disorder obviously enters the averaged $\chi$'s, it is
important to notice that in Eq. (\ref{finalkernel}) the kernel splits
in a part only involving ordered AF couplings and the generic
$\overline\chi_{ij}- \overline\chi_{i0}\overline\chi_{0j}/\overline\chi_{00}$
combination, and a term proportional to $J_D^2$ and to the
purely local susceptibility $\overline\chi_{ii}$. This decomposition
of the kernel into an AF term involving non-local correlations and a 
D term with only local correlations is one of the noticeable results
of this work. Therefore, as far as disorder is concerned, 
any lattice type becomes equivalent to
a Bethe lattice: In the latter case the topology of the lattice allows
for retraceable paths only, while in the former the statistical properties
of the $ J^{D}$'s themselves select the same type of paths.
Therefore it is not surprising that, as in Bethe lattices, 
in the part of the kernel proportional to the disordered interaction 
$\langle 
{\bf S}_{i}(\tau)
{\bf S}_{i}(\tau ')
\rangle^{cavity} = \langle 
{\bf S}_{i}(\tau)
{\bf S}_{i}(\tau ')
\rangle $  for any lattice type.

\subsubsection{Spin self-energy and self-consistent procedure}
In the large-$z$ limit and in the paramagnetic (spin- and
replica-symmetric) phase the local effective action (\ref{complicatedaction})
is cast in the simpler form
\begin{equation}
\label{simpleaction}
{\cal A}
=
-\frac{1}{2}
\int_{0}^{\beta}d\tau\int_{0}^{\beta}d\tau '
{\bf S}(\tau){\bf S}(\tau ')
{\cal K}(\tau-\tau ')~.
\end{equation}
The standard EDMFT procedure is to solve this
impurity problem with a kernel ${\cal K}$
self-consistently depending on the spin ${\bf S}\equiv{\bf S}_{0}^{\alpha}$
via the local susceptibility $\chi_{loc}\equiv {\overline{\chi}}_{ii}$ and an
(approximate) evaluation of ${\overline{\chi}}_{ij}$ entering
Eq. (\ref{completekernel}). Here we emphasize again that,
while the functional dependence of
${\cal K}_{AF}(\omega_n)$ from the spin susceptibilities
(cf. Eq. (\ref{completekernel})) depends on the dimensionality of the
AF spin fluctuations, the disorder part always yields
${\cal K}_D(\omega_n)=J_D^2\chi_{loc}(\omega_n)$.

To solve the impurity problem of Eq. (\ref{simpleaction}) is usually the
difficult step of the EDMFT procedure. In the present paper we
adopt a Quantum Monte Carlo (QMC) technique 
(see subsection II.B) to obtain the local susceptibility
$\chi_{loc}(\omega_n)$.
The next step is then to suitably modify the kernel of Eq. (\ref{simpleaction})
to iterate the self-consistency procedure. To this purpose we customarily
introduce a spin-fluctuation self-energy  $\Pi_{0}^{-1}(\omega_n)$, which, 
owing to the large coordination of the original lattice,
is taken to be momentum independent.
Then in momentum space the non-local susceptibility $\chi_{ij}$
can be written as (see Appendix B) 
\begin{equation}
\label{momentumsuscept}
\chi({\bf q},\omega_n)=\left[ \Pi_{0}^{-1}(\omega_n)+J({\bf q})\right]^{-1}~.
\end{equation}
The calculations reported in Appendix C of Ref. \onlinecite{smithsi}
can easily be generalized (see Appendix B) to the present disordered case, to obtain
the self-energy $\Pi_{0}^{-1}(\omega)$ in terms of the local susceptibility
and  the Fourier-transformed kernel
\begin{equation}
\Pi_{0}^{-1}(\omega_n)={\cal
  K}_{AF}(\omega_n)+\frac{1}{\chi_{loc}(\omega_n)}~. 
\end{equation}
The relation for the averaged local susceptibility 
\begin{equation}
\label{selfcons}
\chi_{loc}(\omega_n)
\equiv \sum_{\qvec} \chi(\qvec,\omega_n)=
\sum_\qvec \frac{1}{\Pi_{0}^{-1}(\omega_n)+J(\qvec)}
\end{equation}
provides another equation allowing to relate completely 
the calculated local susceptibility and
the bosonic bath, allowing to determine from $\chi_{loc}(\omega)$
a new ${\cal K}(\omega_n)$. The procedure is then iterated
until self-consistency is reached.

\subsection{The quantum Monte Carlo approach}

Within our EDMFT scheme, we are faced with the problem of solving
the model in Eq. (\ref{simpleaction}) with an impurity quantum spin coupled to
itself via a retarded interaction kernel representing the surrounding
medium of the other spins. As announced above, we here adopt a numerical
QMC technique, starting from a Hubbard-Stratonovich
transformation of the local partition function
\begin{eqnarray}
Z_{loc}&=&\int {\cal D} {\bf \eta} 
Tr {\cal T} \exp \left[ \int_0^\beta d\tau {\bf \eta}(\tau) \cdot
{\bf S}(\tau) \right]\nonumber \\
&\times&
\exp \left[ -\frac{1}{2}
\int_0^\beta\int_0^\beta d\tau d\tau' {\cal K}^{-1}(\tau, \tau')
{\bf \eta}(\tau) \cdot {\bf \eta}(\tau') \right]~, 
\nonumber \\
&&
\label{zloc}
\end{eqnarray}
where ${\cal T}$ is the imaginary time ordering and $Tr$ denotes a path
integral, taken with respect to the spin degrees of freedom 
${\bf S}(\tau)$. 
This represents the partition function of a spin coupled to an
effective imaginary-time-dependent random magnetic field with
a Gaussian distribution. Notice that within the functional-integral formalism,
the Hubbard-Stratonovich field ${\eta}$ is a bosonic field. This 
relates previous analyses of Bose-Kondo models \cite{smithsiEPL,sengupta}
to our EDMFT analysis of the magnetic model, where the spin
degrees of freedom of surrounding medium are represented by a simpler bosonic bath.

Following Ref.\onlinecite{daniel-marcelo},
to implement the QMC algorithm, we discretize the 
imaginary-time axis into $L$ small time slices so that the time-ordered 
exponential in the trace of Eq. (\ref{zloc}) can be written as the
product of $L$ $2\times 2$ matrices using Trotter's formula.
In our calculations we take the inverse temperature ${J_0}\beta\le 120$
and $L\le 256$ while keeping ${J_0}\Delta \tau ={J_0}\beta /L\le 0.25$,
where ${J_0}$ is a natural magnetic energy scale:
In the 3D case we identify this scale with $J_\times$ (see Sect. IV.A), while in the 2D case
we use the notation ${\hat J}$ (see Sect. IV.B).

\section{Transition criteria and phase diagrams}

\subsection{AF instability criterium}
The establishment of an AF long-range order through a second-order
transition is naturally described within the EDMFT formalism via a divergent
momentum-dependent spin susceptibility 

\beqa
&\chi &({\bf q} \to {\bf Q}_{AF}, \omega=0; T_N) \nonumber \\
&=& \left[\Pi_{0}^{-1}(\omega=0;T_N)+J({\bf Q}_{AF})\right]^{-1}\to \infty
\eeqa
for some specific AF wavevector and a N\'eel temperature 
$T_N$ \cite{smithsi,sinature}.
This would occur when the Fourier transformed magnetic coupling reaches 
an extremal
value, which, e.g., on a hypercubic lattice in $D=z/2$ dimensions, 
yields $J( {\bf Q}_{AF})=2\frac{J}{\sqrt{z}} \sum_{\mu=1}^{D}\cos (Q_{AF})
=-J \sqrt{z}$. 

However, the cumulant expansion and the related
rescaling of $\tilde{J}_{ij}= J_{ij}/\sqrt{z}$, discussed in the previous Section and
needed to keep the dynamics of the quantum spins, would locate
the energy of this ground state very far ($E_0 \sim -\sqrt{z}J$)
from the other states forming the ``body'' of the 
spectrum. Therefore, having in mind that after all
real systems occur only up to three dimensions,
one is naturally led to use a bounded band (with semielliptic and 
rectangular densities 
of states to mimic the three-dimensional and the two-dimensional cases 
respectively). 
In this case the divergence of the spin susceptibility is no longer 
related to 
some specific wavevector corresponding to an easily recognizable space 
arrangement
of the spins, but is rather reached when the border of the band reaches some
specific value. Here, after a suitable rescaling of the energy units, we set 
this
condition to $-J(Q_{AF})\equiv -J =\varepsilon_{min}$ (all the $z$ factors
have been eliminated via the rescaling). Therefore, the condition
for AF order reads 
\begin{equation}
\Pi_{0}^{-1}(\omega =0;T_N)=J~.
\label{AFcondition}
\end{equation}

\subsection{Spin-glass instability criterium}
On the other hand, one can show that the condition
for the instability of the paramagnetic phase toward the formation
of a SG phase is given by
\beqa
1&=&J_D^2\sum_{\qvec} \chi^2(\qvec,\omega_n=0)=
J_D^2\sum_\qvec \frac{1}{\left[\Pi_{0}^{-1}(0)+J(\qvec)\right]^2} \nonumber \\
&=&
J_D^2 \int_{-J}^{J}d \varepsilon 
\frac{\rho(\varepsilon )}{\left[\Pi_{0}^{-1}(0,T_{SG})+\varepsilon \right]^2}
=-J_D^2 \frac{\partial \chi_{loc}(0)}{\partial \Pi_{0}^{-1}} ~,\nonumber \\
&&
\label{SGcondition}
\eeqa
where the spin-excitation density of states $\rho(\varepsilon)\equiv 
\delta(\varepsilon-J(\qvec))$ has been introduced to calculate the momentum sum.
In the particular case when the ordered AF coupling is absent
the action is made local by averaging over the disorder
irrespective of the EDMFT expansion [cf. Section II and Eq. (6)
in Ref. \onlinecite{bray}] and the
condition (\ref{SGcondition}) reduces to $1=J_D^2 \chi_{loc}^2=J_D\chi_{loc}$
\cite{notaJAF0}
This condition was established long ago \cite{bray} for the determination of 
the SG instability for a quantum Sherrington-Kirkpatrick 
model. 

\subsection{Dimensional analysis}
We now aim to establish which one of conditions (\ref{AFcondition}) and 
(\ref{SGcondition}) is realized first depending on the
dimensionality $D$. This latter determines the specific form of
the bare density of states of the spins, which, near the (lower) band edge 
$\varepsilon_{min}=-J$ behaves like
\begin{equation}
\rho(\varepsilon) \propto \left(J+\varepsilon\right)^\frac{D-2}{2}~.
\end{equation}

 To proceed we assume that 
the condition (\ref{AFcondition}) is realized first and we replace
$\Pi_{0}^{-1}$ by $J$ inside Eq.(\ref{SGcondition}) 
\begin{equation}
1 = J_D^2\int_{-J}^J d\epsilon
 \frac{\rho(\epsilon)}{\left(\epsilon+J\right)^2}~.
        \label{newSGcond}
\end{equation}
The convergence of the integral is determined by the band-edge behavior 
at $(\epsilon \approx -J)$.
For $D \le 4$ the integral diverges indicating that, for any finite
value of $J_D$, the condition Eq.(\ref{SGcondition}) is
satisfied before $T$ is lowered to its N\'eel value when $\Pi_{0}^{-1}(T_N)=J$.
In this case the system always becomes a SG
before forming any static AF order. We expect that,
at least for small $J_D$'s, the AF order and a 
replica-symmetry breaking order coexist. It should also be considered
that first-order transitions might occur driving the system
AF or SG ordered without passing through the instabilities
determined by the above conditions \cite{olddaniel}. To investigate and settle 
all these issues,
however, an EDMFT analysis in the presence of
finite AF and/or SG order parameters should be carried out.
This is technically challenging and beyond the scope
of our work. On the other hand, for $D > 4$, the integral in 
Eq. (\ref{newSGcond}) converges. In this case a minimum value of
$J_D$ has to be present, above which a SG order is established
before an AF order parameter is formed. 
Therefore, while for $D \le 4$ even a small amount of disorder 
is enough to produce a SG
phase, for $D>4$ the pure AF order is stable and a sufficiently large 
amount of disorder is required to break the replica symmetry. \cite{noteharris}
\begin{figure}
\includegraphics[angle=-90,scale=0.5]{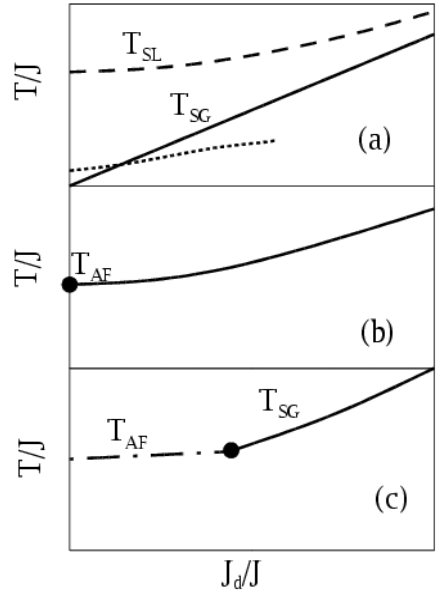}
\caption{Schematic phase diagrams of the disordered AF quantum Heisenberg 
model in (a) $D=2$, (b) $D=3$, and (c) $D>4$. The dashed line $T_{SL}$ is
a crossover temperature below which spin-liquid correlation appear. 
$T_{SG}$ and $T_{AF}$ are the spin-glass and the AF transition
temperatures respectively. The dot schematically indicates the 
position at which the SG temperature becomes larger than the AF one.
The dotted line indicates the ``causal instability'' line occurring in 2D
(see text).}
\label{fig.1}
\end{figure}

Fig. 1 schematically depicts the phase diagrams for
our model in two, three, and above four dimensions.
In two dimensions [Fig. 1(a)], the AF 
long-range order never occurs \cite{notemermin} and
AF correlations only manifest themselves with the
formation of a spin-liquid phase with a reduced
spin susceptibility (with respect to the Curie 
form valid at temperatures above $J$) and with critical
spin dynamics. This behavior arises below a
crossover temperature $T_{SL}$, which has been
numerically determined via the QMC analysis reported
in Section V. This will also allow for the quantitative
determination of the SG transition temperature
$T_{SG}$ and of the spin-liquid instability temperature
$T^*$. This latter marks the temperature below which
the spin self-energy $\Pi_{0}^{-1}(\omega)$ as a function of real frequencies
has a positive imaginary part $\Pi_{0}^{-1}(\omega)''>0$ thereby violating the
causality condition $\omega \Pi_{0}^{-1}(\omega)''<0$. \cite{BGG,haule}

In three dimensions Fig. 1(b) shows that, as soon as a disordered
three-dimensional magnetic coupling $J_D$ is turned on, the
system undergoes a second-order transition to a SG state.
This occurs below a transition temperature $T_{SG}$, which,
according to our criterion above, is always larger than
the N\'eel temperature $T_N$ (in particular it is also
larger than the AF temperature in the absence of disorder $T_N^0$.
In the opposite limit of $J_D \gg J$ one naturally recovers the 
transition temperature for the ``pure'' quantum
Sherrington-Kirkpatrick model \cite{bray,goldschmidt,daniel-marcelo}
and the $T_{SG}/J$ curve 
smoothly approaches the straight line $T_{SG}=0.142 J_D$. 

We finally notice that the qualitative picture reported
in Fig. 1(c) for the case above four dimensions, is
in qualitative agreement with the mean-field calculations
of Refs. \onlinecite{FMSG}, where the competition between
a SG and a ferromagnetic phase was studied.

\section{Static properties}

%
%
\subsection{The three-dimensional case}

When the AF coupling is assumed to act in three dimensions,
the spin-wave spectral density $\rho(\epsilon)$ takes a 
semielliptic form 
\begin{equation}
\rho_{3D}(\epsilon)=\frac{2}{\pi J^2}\sqrt{J^2-\epsilon^2}~.
\label{3Ddensity}
\end{equation}
In this case the integral in Eq. (\ref{selfcons})
can be analytically performed to show that 
this equation is identically satisfied for 
\beq
\Pi_0^{-1}(\omega_n)=J_{\times}^2\chi_{loc}(\omega_n)~,
\label{self-cons3D}
\eeq
with 
\beq
J_{\times}^2 \equiv J^2/4+J_D^2. 
\eeq
Therefore in three
dimensions the problem with mixed AF and disordered 
coupling becomes formally equivalent to a model with
a rescaled coupling $J_{\times}$ only \cite{note3D}.

\subsubsection{Phase diagram}
Starting from the free-spin high-temperature regime, upon
decreasing $T$, the magnetic correlations start to reduce
the magnetic susceptibility. We therefore introduce a
crossover temperature $T_{\times}$, which is arbitrarily
defined as the  temperature at which 
the local magnetic susceptibility is reduced
by ten per cent with respect to the 
free moment Curie law
$4T_{\times}\chi_{loc}(T_{\times})=0.9$.  From the numerical calculations, we find 
$T_{\times}\approx 0.75J_{\times}$. 
Since we decided to use the AF magnetic coupling 
 $J$ as the energy unit, we recast  this 
relation as
\begin{equation}
 \frac{T_{\times}}{J}
\approx 0.75\sqrt{\frac{1}{4}+\left( \frac{J_D}{J}\right)^2 }~,
\end{equation}
where the disordered part of the magnetic interaction
is made explicit.

\subsubsection{Spin-glass temperature}
Using the three-dimensional density of states Eq.~(\ref{3Ddensity}), 
the SG criterion Eq.~(\ref{SGcondition}) is
\begin{equation}
1=
\frac{2}{\pi}
\frac{J_{D}^{2}}{J^2}
\int_{-J}^{J}
\frac{\sqrt{1-\epsilon^2}}{\left[ {\Pi_{0}^{-1}(\omega=0,T_{SG})+\epsilon}\right]^2}
d\epsilon ~.
\end{equation}
Using the relation 
\begin{equation}
\int_{-1}^{1}
\frac{\sqrt{1-x^2}}{\left[ {y+x}\right]^2}dx
=
\frac{\pi}{\sqrt{y^2-1}}\left[ {y -\sqrt{y^2-1}}\right]~,
\end{equation}
the SG criterion can be written as
\begin{equation}
\sqrt{\left({\frac{\Pi_{0}^{-1}}{J}}\right)^2 -1}
=2\frac{J_{D}^{2}}{J^2}
\left[ {\frac{\Pi_{0}^{-1}}{J}-\sqrt{\left({\frac{\Pi_{0}^{-1}}{J}}\right)^2 -1}
}\right]~.
\end{equation}
Finally, this criterion can be cast into the form 
\begin{equation}
\left| {\frac{\Pi_{0}^{-1}}{J}}\right|
=
\frac{1+2(J_D/J)^2}{\sqrt{1+4(J_D/J)^2}}~.
\label{self-encrit}
\end{equation}
It is worth emphasizing again that the self-consistent
value of  $\Pi_{0}^{-1}(T_{SG})$ to be used in the above relations is
the same as in the non-disordered system. Therefore,
to determine the 3D phase diagram as a function of 
temperature and disordered coupling $J_D$, we self-consistently
calculated with QMC the spin self-energy at different temperatures
for a unit value of $ J_\times=J=1$. The results are reported in
the inset of Fig. 2. Then, for each value of $J_D/J$,
the SG instability temperature is obtained
by equating $\Pi_{0}^{-1}(T=T_{SG})$ with the r.h.s. 
of Eq.(\ref{self-encrit}). 
While Fig. 1 is a sketch comparing 
the phase diagrams of our model in various dimensions, 
Fig. 2 shows the phase diagram in three dimensions
as numerically determined from the procedure
outlined above.

It is worth noticing that in the three-dimensional case
the regime with strong spin correlations reducing the 
local susceptibility is severely reduced by the 
occurrence of the SG phase. This is to be contrasted
with the substantial spin-liquid regime found in two dimensions
(see next Section)
\begin{figure}
\includegraphics[angle=-90,scale=0.32]{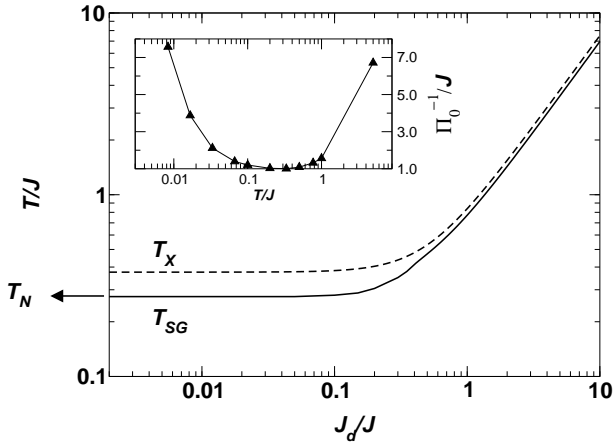}
\caption{Phase diagram of the DQAFH model in three dimensions as obtained from
the QMC calculations at fixed (unit) value of ${J_\times}$. 
 The solid line represents the SG transition temperature $T_{SG}$
as numerically obtained from the criterium of Eqs. (\ref{SGcondition})
and Eq. (\ref{self-encrit}).
The dashed line is the crossover $T_\times$ (see text). Inset: zero-disorder
spin self-energy $\Pi_{0}^{-1}$ as a function of temperature. The 
antiferromagnetic instability
occurs at $T\approx 0.28J$, when $\Pi_{0}^{-1}=J$. }
\label{fig.2}
\end{figure}
\begin{figure}
\includegraphics[angle=-90,scale=0.35]{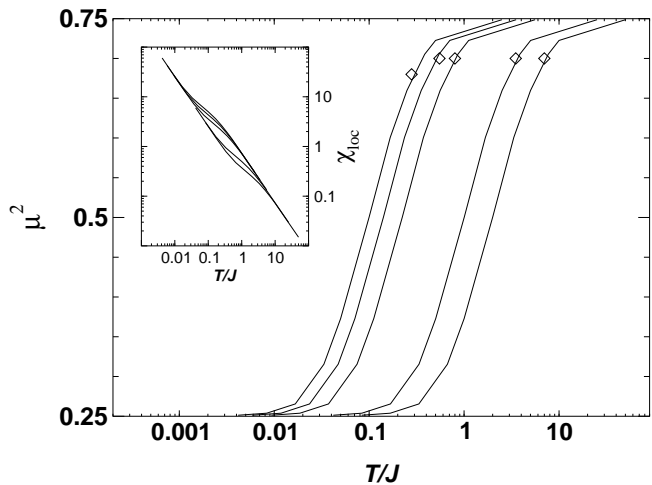}
\caption{Temperature dependence of the square effective moment
$\mu^2(T)=T\chi_{loc}(T)$ in 3D 
at fixed (unit) value of $J_\times$
and various values of disorder:
$J_D/J=0, \, 0.5, \, 1., \, 5, \, 10$ from left to right. 
The high-temperature limiting value is 
$S(S+1)=3/4$, while the low-temperature limit is $1/4$. 
The diamonds show the square moment at the corresponding $T_{SG}$. 
Inset: Temperature dependence of the local spin susceptibility
for $J_D/J=0, \, 0.5, \, 1, \, 5,\, 10$ from top to bottom. }
\label{fig.3}
\end{figure}

\subsubsection{ Local-moment formation}
The temperature dependence of the static local susceptibility
is shown in the inset of Fig.~3 for various values of the disorder $J_D/J$
at a fixed value of $J_{\times}$.
Again, by exploiting the fact that in the three-dimensional case
the disordered coupling can be ``hidden'' inside the effective
coupling $J_{\times}$, we first worked in units of $J_{\times}$, and 
calculated once the universal curve 
$J_{\times}\chi_{loc}(T,J_D,J)=f(T/J_{\times})$). Then we expressed
$J_{\times}$ in terms of $J_D/J$ to find the different 
curves at various values of disorder
in the (meta)stable paramagnetic region, both above and below 
$T_{SG}$. It is quite evident that the susceptibility curves display 
both a high-temperature and a low-temperature Curie-like behavior 
with a constant (unity in log-log plot) universal slope. This
is an indication of local moment formation, as shown in the
main frame of Fig.~3. Here the temperature dependence of the square local moments
$\mu^2(T)=T\chi_{loc}(T)$ is displayed, showing that 
$\mu(T)^2\approx 3/4$ at high temperature, while $\mu(T)^2\approx 1/4$
at low temperature. It can also be noticed that the disordered coupling
strongly favors the formation of static low-T local moments.
However, as shown by the diamonds in Fig. 3, the SG
transition already occurs at much higher temperatures
when the local moments are still close to their high-temperature
value $3/4$.

%
%
\subsection{The two-dimensional case}

We also consider here the case of two-dimensional AF fluctuations
having a spectral density of the form
$
\rho_{2D}(\epsilon)=\frac{1}{2J}
$
for $-J\le \epsilon \le J$. Contrary to the above 3D case, 
the effect of disorder can no longer be included in 
an effective coupling $J_\times$ together with the AF $J$.
Therefore, for any value of $J_D$ one has to recalculate
the self-consistent quantities $\chi_{loc},\,\, {\cal K}$.
These quantities are related by the selfconsistency
Eq.(\ref{selfcons}), which in two dimensions,
reads
\beq
\Pi_{0}^{-1}(\omega, T)=J \coth \left[ J\chi_{loc}(\omega,T)\right]
\label{selfen2D}~.
\eeq

\subsubsection{Phase diagram}
The results of our EDMFT calculations are summarized in the numerical phase diagram
of Fig. 5. 
\begin{figure}
\vspace {1truecm}
\includegraphics[angle=-90,scale=0.35]{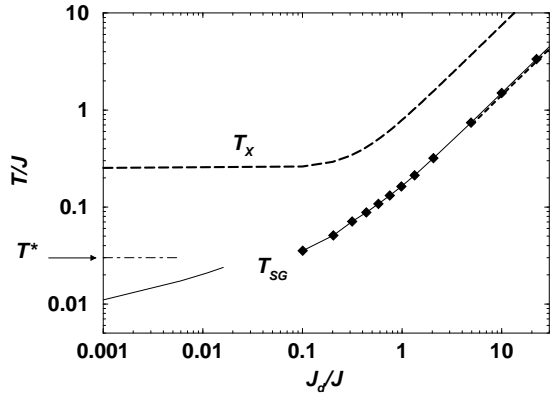}
\caption{Phase diagram of the DQAFH model in two dimensions as obtained from
the QMC calculations. The dashed line represents the analytic
estimation of the crossover temperature $T_\times$ 
the high-temperature free-spin regime to the intermediate-temperature 
spin-liquid regime as obtained from
the high-temperature expansion (see text). The filled diamonds
represent the SG transition temperature $T_{SG}$
as numerically obtained from the criterium of Eq. (\ref{SGcondition}).
The heavy dashed line represents the asymptotic ($J_D\gg J$ transition line for
the quantum Sherrington-Kirkpatrick model. The dot-dashed line
corresponds to the temperature $T^*$ of the
causality violation occurring at zero disorder, while the
thin solid line is obtained from the low-temperature estimation of $T_{SG}$.}
\label{fig4sg}
\end{figure}
Here the presence of various regions
is apparent, where the system displays different physical 
properties. First of all one should notice a crossover
line separating the free-spin high temperature region
from a region at intermediate temperatures and low disorder,
where the spin system has non-trivial correlations. 
This crossover temperature has been determined from 
a high-temperature expansion of Eq. (\ref{selfen2D}),
where the explicit form of $\Pi_{0}^{-1}$ in 2D has been inserted.
Specifically, by assuming that $J\chi_{loc}$ is small at
high temperatures, one obtains that 
\beq
\frac{T_{\times}}{J} =A \sqrt{1+ 9\left({\frac{J_D}{J}}\right)^2}~.
\eeq
The prefactor $A$ is then numerically determined, by imposing that the
spin susceptibility is reduced by ten percent from its high-energy
Curie law. This yields $A\approx 0.25$, for which one obtains the 
dashed line in Fig. 4. Below this line and
before the SG phase takes place at low temperature, the spins
form a correlated ``liquid'' with the static and dynamic
local susceptibilities showing anomalous power-law behavior.
This behavior, at intermediate temperatures where the effects
of $J_D$ are not significant, is quite similar to the case
of the pure two-dimensional system described in Ref. \onlinecite{BGG}.
Specifically, one finds over an order of magnitude in T, where the
static local spin susceptibility displays the quantum-critical
power-law behavior $\chi_{loc}(T) \approx J^{-1}(J/T)^{2/3}$.
Related critical dynamical properties are found in the same
region (see below). However, upon lowering the temperature,
the paramagnetic phase is characterized by the gradual formation
of local moments, whose temperature dependence for various values
of the disorder is reported in Fig. 5.
\begin{figure}
\vspace {1truecm}
\includegraphics[angle=-90,scale=0.35]{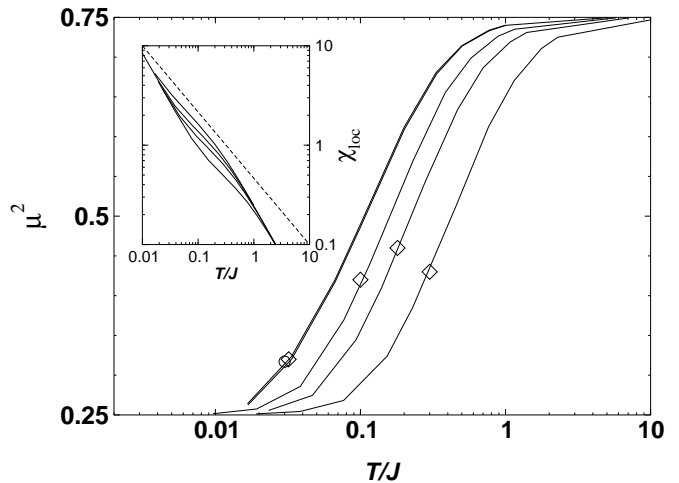}
\caption{Temperature dependence of the square effective moment
$\mu^2(T)=T\chi_{loc}(T)$ in 2D
at various values of disorder:
$J_D/J=0, \, 0.10, \, 0.58, \, 0.98, \, 2.06$ from left to right. 
The high-temperature limiting value is 
$S(S+1)=3/4$, while the low-temperature limit is $1/4$. 
 The circle in the $J_D/J=0$ curve shows the square moment at $T=T^*\approx 0.03J$.
The diamonds show the square moment at the corresponding $T_{SG}$.
Inset: Temperature dependence of the local spin susceptibility
for $J_D/J=0, \, 0.58, \, 0.98, \, 2.06$ from top to bottom. The dashed line
is a guide for the eyes corresponding to a power-law behavior with 
an exponent $2/3$ to illustrate the intermediate spin-liquid behavior.}
\label{fig5sg}
\end{figure}

\subsubsection{ Local-moment formation}
The fact that disorder favors the formation of local
moments at progressively larger temperatures, is made apparent
by the flattening of the curves to the classical value 1/4.
It is worth noticing that the tendency to form a finite
local moment at low temperature is also present in the
absence of disorder ($J_D=0$). This tendency was
also just visible in our previous work\cite{BGG},
in our QMC calculations at $T=J/100$. However, 
it was impossible to confirm 
this numerical hint 
by well-converged QMC calculations at lower temperatures
\cite{notenomomentum}. On the other hand, the static moment
formation becomes quite clear in the presence of disorder.

\subsubsection{Spin-glass temperature}
So far we have discussed the properties of the paramagnetic
phase disregarding the possible occurrence of a SG 
instability as determined by Eq. (\ref{SGcondition}).         
In the present two-dimensional case,
the relation Eq. (\ref{SGcondition}) characterizing the
SG temperature can be simplified to 
\beq
\Pi_{0}^{-1}(\omega=0, T_{SG})=\sqrt{J_{D}^{2}+J^{2}} \equiv \hat J~,
\eeq
with $\Pi_{0}^{-1}(\omega,T)$ given by Eq. (\ref{selfen2D})
In the weak disorder limit $J_D<<J$, we approximate $\chi_{loc}$ 
by its zero disorder expression 
$J\chi_{loc}(\omega=0, T<<J)\approx 0.338(J/T)^{2/3}$
(see Fig. 1(b) of Ref. \onlinecite{BGG}).
This gives the asymptotic behavior in the limit $J_D/J\rightarrow 0$
\begin{equation}
T_{SG}\approx
J\left[ \frac{0.338}{\ln\frac{J}{J_D}}\right]^{3/2}~,
\label{tsg2d}
\end{equation}
showing that a significant $T_{SG}$ can be obtained even for a small 
disorder. 
The numerical data seem indeed to be smoothly extended to this
low-disorder behavior (see Fig. 4).

\section{Dynamical properties}

\subsection{Dynamical properties in 3D}
\begin{figure}
\includegraphics[angle=-90,width=8cm]{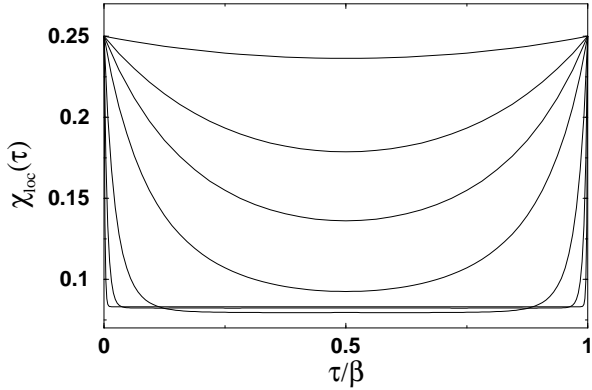}
\vspace {-1truecm}
\caption{Imaginary-time dependence of the local susceptibility 
in three dimensions at different values of the inverse temperature 
$\beta J_\times= 1, \, 3, \, 5, \, 10, \, 30, \,60, \, 120$.}
\label{fig6sg}
\end{figure}

From our QMC analysis we also obtained dynamical quantities
in imaginary time. 
Fig. \ref{fig6sg} displays the imaginary-time dependence of the local spin 
susceptibility $\chi_{loc}(\tau)$ {\it in the paramagnetic phase}. 
\cite{notaplot}
Indeed, although
the system rapidly passes from a paramagnetic high temperature phase to 
a low-temperature SG phase (see Fig. 2), we follow the paramagnetic
phase down to low temperatures to allow a comparison with the
two-dimensional case, where the critical SG temperature
is much lower. In this way we can investigate the role of dimensionality 
in determining the spin dynamics. 

In  $\chi_{loc}(\tau)$ one can recognize the different 
temperature regimes found in the static quantities and 
explored in detail in the previous study of the purely disordered
case in Ref. \onlinecite{daniel-marcelo}. This is quite obvious
since, as already discussed for the static properties,
our system in three dimensions is equivalent to a purely
disordered Heisenberg model with an effective magnetic
coupling [see Eq.(\ref{self-cons3D})]. 
This equivalence turns out to be quite important, since
the dynamical analysis involves a delicate procedure of
extracting real-frequency information from data obtained
in imaginary time or frequencies. This is usually a difficult
task, but it is greatly simplified by the knowledge of 
well-established limits at low and high temperatures for
the dissipative part of the local response $\chi_{loc}(\omega)$.
This allows the construction of an interpolating function that
successfully matches the numerical QMC data.
Specifically one finds
a high-temperature regime, where the spin dynamic absorption
produces a Gaussian quasi-elastic peak 
\beqa
\frac{\chi_{loc}''(\omega)}{\pi \omega}&=&
\frac{\beta S(S+1)}{3}\left[1-S(S+1)\frac{(\beta {J_\times})^2}{18}\right] \nonumber \\
&\times& \frac{e^{-1/2(\omega/\omega_L)^2}}{\sqrt{2\pi\omega_L^2}}~,
\label{highTchiomega}
\eeqa
with a single characteristic relaxation frequency $\omega_L$ given
by
$\omega_L^2=2{J_\times}^2S(S+1)/2[1-S(S+1)(\beta {J_\times})^2/18]$
In imaginary-time this corresponds to
\beqa
\chi_{loc}^L(\tau)
&=& \frac{S(S+1)}{3}e^{-\frac{1}{2}\left(\frac{\beta\omega_L}{2}\right)^2
\left[1-\left(1-\frac{2\tau}{\beta}\right)^2\right]} \nonumber \\
&\equiv&\frac{S(S+1)}{3}\Phi_L(\tau) ~.
\label{highTchitau}
\eeqa
On the other hand, at low temperatures the curves in Fig. 6 clearly
display two regimes: At long times the curve flattens indicating
the formation of nearly static moments, while at 
short times a rapid variation (decrease)
of $\chi_{loc}(\tau)$ indicates that some spectral weight is shifted
at high frequencies. From an analysis similar to the one
of Ref. \onlinecite{daniel-marcelo} one can see that this 
leads to dissipation from transverse spin excitations 
\beqa
&&\frac{\chi_{loc}^T(\omega)}{\pi \omega} =
\frac{S}{2}\left[\frac{\beta}{2\omega_T}\right]^{3/2} 
\frac{\omega \tanh(\beta \omega/2)}{\sqrt{2\pi}(1+\beta\omega_T/2)}
\nonumber \\
& \times & \left[
\exp\left[-\frac{\beta (\omega-\omega_T)^2}{4\omega_T} \right] +
\exp\left[-\frac{\beta (\omega+\omega_T)^2 }{4\omega_T}\right]
\right]~, \nonumber \\
&&
\label{lowTchiomega}
\eeqa
at frequencies $\omega_T \sim {J_\times}^2/T$.
In imaginary time this gives
\beqa
\chi_{loc}^T(\tau)
&=& \frac{S(S+1)}{3}\frac{1+\beta\omega_T/2(1-2\tau/\beta)^2}{1+\beta\omega_T/2} \nonumber \\
&\times&
\exp\left[{-\frac{\beta\omega_T}{4}\left[1-\left(1-\frac{2\tau}{\beta}\right)^2\right]}
\right] \nonumber \\
&\equiv & \frac{S(S+1)}{3}\Phi_T(\tau)~.
\label{lowTchitau}
\eeqa

The longitudinal fluctuations instead still produce a quasi-elastic absorption
peak of the form of Eq. (\ref{highTchiomega}), 
but with a reduced frequency $\omega_L\propto T$ and a prefactor
$\beta S^2/3$ replacing $\beta S(S+1)/3$. This reduction of the
effective Curie constant is natural  since the formation
of static moments must coexist with the high-energy absorption. 

The known behavior in the high- and low-temperature regimes
allows to construct an interpolation function 
\beq
\chi_{loc} =\frac{S(S+1)}{3}\left[ p\Phi_L(\tau) + (1-p)\Phi_T(\tau) \right]~.
\label{interpolate3D}
\eeq
Obviously the temperature variation of the relaxation function
$F(\omega)\equiv \chi''(\omega)/(\pi \chi_{loc}(T) \omega)$
is quite similar to the one shown in Fig. 2 of Ref. \onlinecite{daniel-marcelo}
provided that $J$ is replaced by ${J_\times}$.

\subsection{Dynamical properties in 2D}
Fig. 7 displays the results of our QMC calculations for the
imaginary-time dependent spin susceptibility. The frames from top to bottom
are for different values of disorder  $J_D=0.1J$, $J_D=J$, and $J_D=10J$.
The various curves are calculated at different 
values of the inverse temperature $\beta \hat J=1,\, 3, \,
5, \, 10, \, 30, \, 60$. Frames (b) and (c) also contain the
curves for $\beta \hat J=120$, which could not be obtained
for the zero disorder case due to non-convergence of the QMC code.
Similarly to the 3D case, the low-temperature curves clearly display
a flattening at large time marking the tendency to form static local moments.
This formation is clearly favored by disorder. The same conclusion
could qualitatively be drawn from  the local susceptibility as a
function of Matsubara imaginary frequencies for various values of
the disorder: At low temperature the more disordered
systems display a visible tendency to form an additional contribution at 
zero frequency, which cannot be obtained from a smooth
extrapolation from the finite-frequency data. This is because the local
moments provide a substantial delta-like contribution to the zero-frequency
susceptibility (however, due to 
the difficulty in reliably extracting this additional contribution, we 
refrain from showing curves for $\chi_{loc}(\omega_n)$).
For the curves of $\chi_{loc}$ as a function of imaginary-time,
we performed an  analysis similar to the one carried out in Subsection
V.A for the three-dimensional case. 

Upon inspecting the two-dimensional phase diagram of Fig. \ref{fig4sg} one sees that
a substantial region is present between $T_\times$ and $T^*$ or $T_{SG}$,
where strong spin correlations should give rise to a non-trivial spin dynamics.
From our previous work on the non-disordered Heisenberg model in two dimensions \cite{BGG},
we learned that a spin-liquid phase arises below $T_\times$. At long imaginary times
this phase has a critical susceptibility of the form
$\chi_{loc}(\tau)\sim [sin(\pi\tau/\beta]^{-1/3}$, which corresponds
to a scaling of the imaginary local susceptibility \cite{BGG} 
\begin{equation}
\label{imchiloc}
\chi''_{loc}(\omega)\sim J^{-1}(\omega/J)^{-\delta} {\cal F}_\delta(\omega/T)~,
\end{equation}
with ${\cal F}_\delta(x)=x^{\delta}\vert \Gamma(\frac{1-\delta}{2} 
+i\frac{x}{2\pi})
\vert^2 \sinh\left(\frac{x}{2} \right)$ and $\delta = 2/3$.
Eq.~(\ref{imchiloc}) gives a power-law divergence at low energy, 
$\chi''_{loc}(T=0,\omega) \sim J^{-\frac{1}{3}}\omega^{-\frac{2}{3}}$ and 
$\chi'_{loc}(T,\omega=0)\sim J^{-\frac{1}{3}}T^{-\frac{2}{3}}$. 

The question then naturally arises, whether or not this critical behavior
can be experimentally observed in magnetic systems with strong
quantum fluctuations, weak disorder, and strong anisotropy in the AF 
fluctuations.

In principle one could answer this question by extending the analysis
of Subsect. V.A and generalizing Eq. (\ref{interpolate3D}) to a form
including an additional spin-liquid contribution 
\beq
\chi_{loc} =\frac{S(S+1)}{3}\left[ p\Phi_L(\tau) + q \Phi_{SL}(\tau)
(1-p-q)\Phi_T(\tau) \right]~.
\label{interpolate2D}
\eeq
with $\Phi_{SL}(\tau)\propto [sin(\pi\tau/\beta]^{-1/3}$.
Starting from the Fourier transform of Eq. (\ref{interpolate2D}), we
tried to use a standard procedure based on Pad\'e approximants to 
analytically continue the Matsubara susceptibility to real frequencies.
Unfortunately, the analytic continuation procedure is so delicate
that reliable results can hardly be extracted from a form like
Eq. (\ref{interpolate2D}). In particular we noticed that small variations
in the relative weights $p,q$ as well as in the other physical parameters
$\omega_L$ and $\omega_T$ lead to substantial variations in the real-frequency
susceptibility. Therefore, while our analysis {\it is fully compatible with} 
the possibility
of intermediate frequency windows with anomalous spin-liquid
exponents, we cannot provide reliable bounds to the observation of these 
regimes.

\begin{figure}
\includegraphics[angle=-90,scale=0.75]{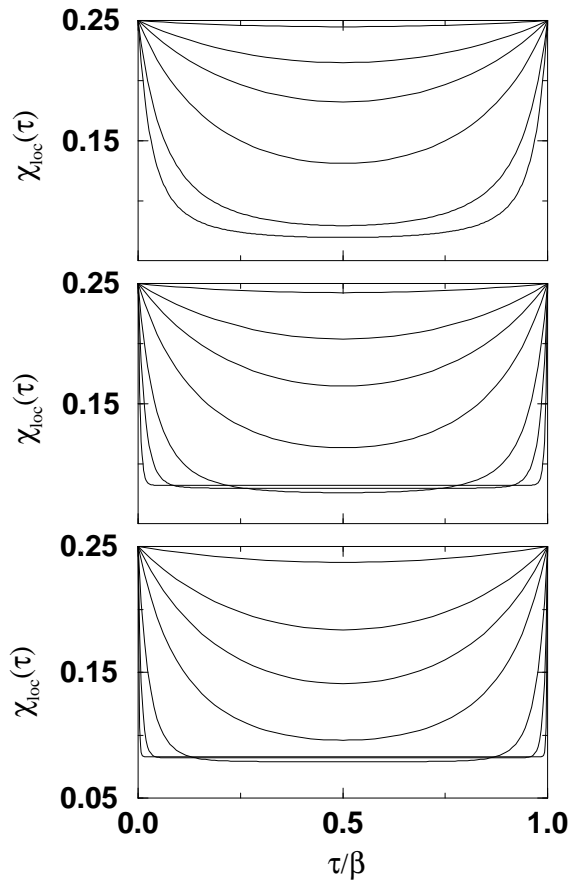}
\vspace {-1.5truecm}
\caption{Imaginary-time dependence of the local susceptibility 
in $D=2$ for
different values of the inverse temperature $\beta \hat J=1,\, 3, \,
5, \, 10, \, 30, \, 60, \, 120$ and three
values of disorder: (a) $J_D=0.1J$, (b) $J_D=J$, and (c) $J_D=10J$.
For the low-disorder case (a) the lowest-temperature data $\beta\hat J=120$
are not available (see text).}
\label{fig7sg}
\end{figure}

\section{Discussion and conclusions}
In this paper we systematically investigated the interplay between 
disorder and fluctuations in a quantum spin-$\frac{1}{2}$ Heisenberg
model. 
Without entering any broken-symmetry phase and neglecting the
possible occurrence of first-order transitions, we investigated
the properties and the instabilities of the paramagnetic phase.
We found that a marked difference exists between the two- and the
threedimensional cases. In 3D the system is ordered at a finite N\'eel
temperature $T_{AF}$ for zero disorder and becomes a SG at 
$T_{SG}>T_{AF}$ as soon as a finite disorder is introduced in the
magnetic coupling. This result is not so surprising because
the AF and the SG phase are not competing 
so that antiferromagnetism can well coexist
with glassiness at least for small disorder. 
According to these indications, in real AF systems
the presence of a small amount of disorder is expected
to lead to weak glassy properties. These, however, can likely
be masked by the dominant (at weak disorder) AF properties.

The two-dimensional case is richer. In the absence of disorder, the spin liquid phase
analyzed in Ref. \onlinecite{BGG} was shown to be unstable
below a temperature $T^*\sim 0.03J$, where the causality condition
 $\omega \Pi_{0}^{-1}(\omega)''<0$
on the self-energy of the dynamical local spin susceptibility starts to be violated.
As discussed in detail in Ref. \onlinecite{rosch} for the $t-J$ model, this
behavior occurs when long ranged space correlations arise at finite
temperature being driven by a divergent
magnetic coherence length $\xi$. In this case in 2D the real part of 
the local susceptibility diverges because the momentum-independent spin self-energy
introduced in the EDMFT approach
is unable to produce anomalous dimensions (and the related  $\eta$ exponent).
Then $\chi_{loc}'(0)\sim \log \xi$ 
(unphysically above the bare Curie value~\cite{falk} $S(S+1)/3T$).
This growth at low (but still finite) temperatures entails too
large values of the imaginary part of $\chi_{loc}''$, which give an unphysical sign of the 
spin self-energy determined by the self-consistency condition Eq. (\ref{selfen2D}).
This occurs at the ``causal instability'' temperature $T^*$. 
Nevertheless, the system may display an anomalous behavior of the spin susceptibility
over broad frequency and temperature ranges before the instability
occurs. Therefore the EDMFT treatment keeps a valuable physical interest
to detect non-trivial spin correlations in low dimensions.
Furthermore, we showed here that the causal instability may be prevented by
a small amount of disorder: As seen in Fig. 4, already at $J_D \approx
(0.05 - 0.1)J $ the SG transition temperature overcomes $T^*$.
This also indicates that EDMFT acquires further interest and validity
to treat those models where physical mechanisms prevent the large
growth of the local susceptibility or induce the transition to a 
stable phase\cite{notatstar}.

A noticeable finding of this paper is also related to the specific
form of the EDMFT kernel obtained in Eq. (\ref{completekernel}).
In particular, as shown in Subsect. II.A and in Appendix B,
the kernel separates into a
AF and a disordered part (although disorder is effectively present
in both terms via the averaged $\chi_{ij}$'s) and moreover the disordered
kernel is build up with the local susceptibility only
\beq
{\cal K}_{D} = (J_D/z)\sum_i{\overline \chi}_{ii}~.
\label{kappaD}
\eeq
 The crucial ingredient 
leading to these two results were the local nature of the disorder
($\langle J_{ij}^{D} J_{kl}^{D}  \rangle_{dis}=\delta_{i,l}\delta_{j,k}J_D^2$)
and the large coordination of the lattice ($z\to \infty$).
It is interesting to notice that  the specific form in Eq.(\ref{kappaD}) 
of the disorder-kernel holds for any lattice type and is clearly reminiscent of the form
obtained in the case of an ordered system on a Bethe lattice\cite{reviewDMFT}.
Formally this arises because ${\cal K}_D$ is only non-vanishing  
for $\langle J_{ij}^{D} J_{kl}^{D}  \rangle_{dis}=\delta_{i,l}\delta_{j,k}J_D^2$,
which also holds in the specific case of the Bethe lattice where only fully
retraceable paths are allowed. One consequence of the generic validity of 
Eq. (\ref{kappaD}) is that the results of 
Bray and Moore in Ref. \onlinecite{bray} are naturally recovered here (both in
two and three dimensions) in the limit of zero AF coupling.
Indeed the (cubic) lattice with infinite-range interaction of Ref. \onlinecite{bray}
becomes equivalent to a lattice with infinite coordination because
the role of the thermodynamic limit ($N\to \infty$, with $N$ number of sites)
is replaced here by the $z\to \infty$ limit.  Since the specific form of the
lattice becomes immaterial for Gaussian (local) disorder, our results
in the $J_D/J\to \infty$ limit where ${\cal K}\approx {\cal K}_D$ recover
both the variational and the QMC results of Refs. \onlinecite{bray,daniel-marcelo}.
In particular, as shown in Sect.IV,  we find the known linear dependence
$T_{SG}=0.142 J_D$ both in 2D and 3D.
Our results, however, also show that in the presence of a uniform
AF coupling the SG transition temperature occurs at values larger than the 
above linear behavior. Physically this occurs because quantum 
and thermal fluctuations are naturally decreased and severely suppressed 
by the occurrence of the AF phase in 3D. In other words, once AF correlations
nearly quenches the spins, a (even small) disorder $J_D$
easily produces a glassy state at finite temperature $T_{SG}$
In 2D this mechanism is still present but weaker, because the AF order does
not establish and the spin fluctuations are only reduced by
the spin-liquid correlations. Then $T_{SG}$ vanishes for $J_D$ tending to zero, but 
slowly because of the logarithmic dependence in Eq. (\ref{tsg2d}).

\acknowledgments
S.B. and M.G. are very sorry to announce that their friend and coworker
Daniel Grempel passed away during the preparation of this manuscript. 
We will miss his inestimate friendship and scientific support. 
We would like to recognize his contribution to this work whilst 
assuming full responsibility for its posthumous publication. 

We thank C. Castellani, L. Cugliandolo, I. Giardina, and
Q. Si  for useful discussions and suggestions.
S.B. acknowledges  financial support from
the FERLIN Program of the European Science Foundation.
M.G. acknowledges financial support from the MIUR PRIN 2005 - prot. 2005022492.

\appendix
\section{Derivation of the local action}

\subsection{The replica trick}
We consider the DQAFH model defined by the Hamiltonian
Eq.~(\ref{hamiltonian}), 
and we consider the effect of disorder within the
replica trick\cite{Replicatrick}. 
The replica trick relies on the assumption that the free energy ${\cal F}$ 
of the system is a self-averaging quantity. 
This means that in the thermodynamic limit the free energy does not depend on
the configuration chosen for the random couplings $J_{ij}^{D}$. 
As a consequence, ${\cal F}$ remains unchanged when it is averaged with
respect to all the possible configurations of $J_{ij}^{D}$, and we find 
\begin{eqnarray}
\beta{\cal F}
&=&
\langle log {\cal Z}(J_{ij}^{D} \rangle_{dis} \nonumber\\
&=&
\lim_{n\rightarrow 0}
\frac{\langle  {\cal Z}(J_{ij}^{D}) \rangle_{dis} -1}{n} \nonumber\\
&=& 
\lim_{n\rightarrow 0}\frac{1}{n}
log\langle {\cal Z}^n(J_{ij}^{D})\rangle_{dis}~, 
\end{eqnarray}
where ${\cal Z}(J_{ij}^{D})$ is the partition function of the system for a
given configuration of $J_{ij}^{D}$, and $\langle\cdots\rangle_{dis}$ denotes
an average with respect to the probability distribution $P(J_{ij}^{D})$. 

One of the replica trick hypotheses consists of assuming that the average 
$\langle {\cal Z}^n(J_{ij}^{D})\rangle_{dis}$ is an analytical function of
$n$. We can thus express this quantity for integer values of $n$, interpreting 
${\cal Z}^n(J_{ij}^{D})$ as being the partition function of 
$n$ replicas of the system which are independent of each other but
characterized by the same distribution of $J_{ij}^{D}$. 
This is formally expressed by the relation 
\begin{equation}
\langle {\cal Z}^{n}(J_{ij}^{D})\rangle_{dis}
=\prod_{i<j}\int_{-\infty}^{+\infty}dJ_{ij}^{D}P(J_{ij}^{D})
Tr\left[ \exp -\sum_{\alpha=1}^{n}H_\alpha\right]~, 
\end{equation}
where $H_\alpha$ is the disordered Heisenberg Hamiltonian 
Eq.~(\ref{hamiltonian}) acting on the replica $\alpha$. 
In Eq.(\ref{zndis}), the trace $Tr$ is applied to all the operators 
${\bf S}_{i}^{\alpha}$ of the replicated Hilbert space. 
Here, the integration with respect to $J_{ij}^{D}$ generates an effective 
action with the following properties: \\
{\it (i)} The corresponding effective model now recovers the translational
symmetry of the underlying lattice. This allows a dynamical mean field (DMFT) 
approach. \\
{\it (ii)} The spins of the same replica are coupled through the
antiferromagnetic exchange $J_{ij}^{AF}$, and an antiferromagnetic (AF) 
instability can occur. \\
{\it (iii)} Some correlations occur between different replicas of the same spin, 
which could give rise to a ground state characterized by a replica symmetry
breaking. This state would describe a SG phase.

\subsection{DMFT and the local partition function}
Following the usual DMFT approach\cite{reviewDMFT}, we consider the
local effective partition function seen from a given site of the lattice, that
we arbitrarily denote $i=0$. 
As a preliminary, we introduce the 'cavity' Hamiltonian $H^{(0)}$, 
corresponding to the Hamiltonian $H$ without the site $i=0$. 
The complete Hamiltonian is related to the cavity one with 
\begin{eqnarray}
H
=
H^{(0)}
+
\sum_{j\neq 0}V_{0j}~, 
\end{eqnarray}
with
\begin{eqnarray}
V_{0j}=
\frac{J_{0j}^{AF}+J_{0j}^{D}}{\sqrt{z}}
\sum_{\alpha=1}^{n}
{\bf S}_{0}^{\alpha}{\bf S}_{j}^{\alpha}~, 
\end{eqnarray}
describing the magnetic interaction between the site 
$i=0$ and its neighbors. The usual static mean field approximation 
consists of replacing this term by its average with respect to the thermal and 
disorder fluctuations. 
In a paramagnetic phase, this contribution would vanish and we would have to go
beyond the static mean field, within the DMFT approach. 
Using a path integral formalism, the partition function of the replicated
system is given by 
\begin{eqnarray}
{\cal Z}^{n}(J_{ij}^{D})
=
Tr {\cal T}\exp \left[-\int_{0}^{\beta}d\tau H(\tau)\right]~, 
\end{eqnarray}
where ${\cal T}$ is the imaginary time ordering and 
the trace $Tr$ denotes a path integral, taken with respect to all the spin degrees of
freedom ${\bf S}_{i}^{\alpha}(\tau)$. 
This is now formally performed in two steps 
$Tr=Tr_0Tr^{(0)}$, by considering first all the degrees of freedom 
concerning the sites $j\neq 0$ ($Tr^{(0)}$), and then computing the
partial trace for the local spin operators on site $i=0$ ($Tr_{0}$). 
We find
\begin{eqnarray}
&{\cal Z}^{n}(J_{ij}^{D})& 
=  \nonumber\\
&Tr_{0}Tr^{(0)}&
{\cal T}\exp\left[-\int_{0}^{\beta}d\tau \left(H^{(0)}(\tau)
-\sum_{j\neq 0}
V_{0j}(\tau)\right)
\right]~. \nonumber\\
&& 
\end{eqnarray}
We average this relation with respect to the random distribution of 
the couplings $J_{ij}^{D}$, assuming that the probability distribution 
$P(J_{ij}^{D})$ has no correlation between two different pairs of sites
$(i,j)$. The averages for the cavity (with $H^{(0)}$) and the local 
(with $V_{0j}$) parts can thus be factorized 
\begin{eqnarray}
&&\langle {\cal Z}^{n}(J_{ij}^{D})\rangle_{dis}={\cal Z}^{(0)}
\nonumber
\\ 
&\times&
\langle {Tr_{0}\langle {
{\cal T} \exp\left[ -\sum_{j\neq 0}\int_{0}^{\beta}d\tau V_{0j}(\tau) \right]
}\rangle^{(0)}}\rangle_{dis}~, 
\label{Relationpartition1}
\end{eqnarray}
where 
\begin{eqnarray}
{\cal Z}^{(0)}
=
\langle
Tr^{(0)}
{\cal T}\exp\left[-\int_{0}^{\beta}d\tau H^{(0)}(\tau)\right]
\rangle_{dis}~, 
\end{eqnarray}
and $\langle\cdots\rangle^{(0)}$ denotes the thermal and disorder 
average with respect to the cavity Hamiltonian $H^{(0)}$. 
This is performed by using a cumulant expansion, using the general formula 
$\langle {\cal T} \exp[A]\rangle^{(0)}={\cal T} \exp[\sum_{p=1}^{+\infty}C_{p}/p!]$, where 
$A=-\sum_{j\neq 0}\int_{0}^{\beta}d\tau V_{0j}(\tau)$ and $C_{p}$ is the 
$p-$order cumulant, which invokes correlation functions between $p$ operators 
${\bf S}_{j}^{\alpha}$. 
The usual static mean field approximation would consist of neglecting 
all the cumulants $p\geq 2$. But, since here we consider the paramagnetic 
phase, all the cumulants $C_{p}$ with odd $p$ vanish. 
Furthermore, in the limit $z\to\infty$ we find $C_{p}\approx z^{1-p/2}$, and 
$\langle {\cal T}\exp[A]\rangle^{(0)}={\cal T}\exp[C_{2}/2]$. 
The DMFT approach, which is exact in the limit of a large coordination 
number $z$, thus consists of approximating the relation 
Eq.~(\ref{Relationpartition1}) with 
\begin{eqnarray}
\langle {\cal Z}^{n}(J_{ij}^{D})\rangle_{dis}
= 
{\cal Z}^{(0)}
\langle {Tr_{0}\langle {
{\cal T}\exp {\frac{C_{2}}{2}}
}\rangle^{(0)}}\rangle_{dis}~, 
\end{eqnarray}
where
\begin{eqnarray}
C_{2}
=
\int_{0}^{\beta}d\tau
\int_{0}^{\beta}d\tau '
&\sum_{\nu\nu'}\sum_{\alpha\alpha'}
S_{0\nu}^{\alpha}(\tau)S_{0\nu'}^{\alpha'}(\tau')& 
\nonumber\\
\times&\sum_{ij}
\frac{J_{0i}J_{j0}}{z}
\langle {
{\cal T} S_{i\nu}^{\alpha}(\tau)S_{j\nu'}^{\alpha'}(\tau')
}\rangle^{(0)}~. &
\nonumber\\
&&
\end{eqnarray}
From this expression Eq. (\ref{complicatedaction}) is then obtained,
which reduces to Eq. (\ref{simpleaction}) in the paramagnetic case.

\section{Disorder average of the action kernel}
\subsection{Preliminaries}
Our treatment of disorder proceeds along the following steps: 
First we fix the disorder configuration assuming a specific
realization of the $J_{ij}^{D}$ distribution. 
Then, for this configuration one can closely follow the same steps as in 
Ref.~\onlinecite{smithsi} to derive the relation between the spin correlation 
functions with and without cavity
\begin{equation}
\label{cavitychi}
\chi_{ij}^{cavity}
=
\chi_{ij} - \chi_{i0}\chi_{0j}/\chi_{00}~.
\end{equation}
Here, we want to establish a similar relation for the disorder-averaged 
susceptibilities $\overline{\chi}_{ij}$ defined by Eq.~(\ref{defchiaverage}). 

In general one can see that the various correlation functions are obtained by
sums of the type 
\begin{eqnarray}
{{\chi}}_{ij}
&=&
\delta_{ij}{\Pi}_{ii}
-
\frac{J_{ij}}{\sqrt{z}}{\Pi}_{ii}{\Pi}_{jj}
\nonumber \\
&-&
\sum_{paths}
\frac{(-1)^{p}J_{ii_1}
J_{i_1 i_2}
\cdots 
J_{i_p j}
}{z^{(p+1)/2}}
{\Pi}_{ii}
{\Pi}_{i_1 i_1}
\cdots
{\Pi}_{i_p i_p}
{\Pi}_{jj}~, 
\nonumber \\
&&
\label{Expansionchi}
\end{eqnarray}
where, to obtain the leading contribution in the large-$z$ expansion,
the sum is taken over non-self-intersecting paths and the local correlations 
${\Pi}$ take into account loop decorations of these paths. 
We will see in the following how these local propagators can be related to a
local self-energy, which will be determined in a self-consistent way. 

The relation (\ref{cavitychi}) for the cavity susceptibility 
is obtained from Eq.~(\ref{Expansionchi}) by using the general 
algebraic identity 
$\sum_{i\to 0}\sum_{0\to j}
=[\sum_{i\to j}-\sum_{i\to j}^{cavity}]\sum_{0\to 0}$, 
where $\sum_{i\to j}$ represents a summation over all the 
direct paths between two sites $i$ and $j$, and $\sum_{i\to j}^{cavity}$ 
excludes all the direct paths through the site $0$. 

\subsection{Disorder averaging of the kernel}
Here we will average the large$-z$ expansion (\ref{Expansionchi}) 
with respect to all the possible distributions of $J_{ij}^{D}$. 
First, we remark that a given bond $l-m$ (and the related $J_{lm}$)
either appears in the product of $J$'s, forming the ``skeleton'' of the
path, or is inside the loops forming one (and just one) of the $\Pi$'s. 
As a consequence, when taking the average over the disorder, each object,
a $J$ or a ${\Pi}$ is separately averaged with respect to the others;
therefore the average over the disorder
distribution $P(J_{ij}^{D})$ factorizes in the average
of each separate ${\Pi}$ and of the product of $J$'s. 
Furthermore one should notice that,
since the paths are non-self-intersecting, the product of $J$'s can only
contain one $J$ per $l-m$ bond. Therefore, owing to the zero average of
the disordered $J^D$, this product vanishes unless it only
contains purely AF ordered couplings. Then 
\begin{eqnarray}
{\overline{\chi}}_{ij}
&=&
\delta_{ij}{\overline \Pi}_{ii}
-
\frac{J^{AF}_{ij}}{\sqrt{z}}{\overline \Pi}_{ii}{\overline \Pi}_{jj}
\nonumber \\
&-&
\sum_{paths}
\frac{(-1)^{p}J^{AF}_{ii_1}
J^{AF}_{i_1 i_2}
\cdots 
J^{AF}_{i_p j}
}{z^{(p+1)/2}}
{\overline \Pi}_{ii}
{\overline \Pi}_{i_1 i_1}
\cdots
{\overline \Pi}_{i_p i_p}
{\overline \Pi}_{jj}~,
\nonumber \\
&&
\label{Expansionchiaverage}
\end{eqnarray}
where the bar indicates disorder-averaged quantities. Then the 
same arguments used in Ref. ~\onlinecite{smithsi} for obtaining the relation 
(\ref{cavitychi}) from the expansion (\ref{Expansionchi}), can now be applied
to the averaged quantities of Eq.~(\ref{Expansionchiaverage}) giving
\begin{equation}
\label{cavitychiaverage}
\overline{\chi}_{ij}^{cavity}
=
\overline{\chi}_{ij} - 
\overline{\chi}_{i0}\overline{\chi}_{0j}/\overline{\chi}_{00}~. 
\end{equation}
This completes the derivation of Eq. (\ref{averagechicavity})
leading to the final expression of the kernel Eq. (\ref{completekernel})

\subsection{Local self-energy}
The expansion (\ref{Expansionchiaverage}) is equivalent to a Bethe-Salpeter equation 
for the average susceptibilities
\begin{equation}
{\overline{\chi}}_{ij}(\omega_{n})
=
\delta_{ij}{\Pi}_{0}(\omega_{n})
-
{\Pi}_{0}(\omega_{n})
\sum_{l}\frac{J_{il}^{AF}}{\sqrt{z}}{\overline{\chi}}_{lj}(\omega_{n})~, 
\label{EquationBS}
\end{equation}
where we restored the explicit Matsubara frequency dependency, and 
the averaged 
${\Pi}_{0}\equiv\overline{\Pi}_{ii}$ is site independent. 

Averaging the distribution of $J_{ij}^{D}$ restores the translational symmetry
of the underlying lattice. We can thus define the ${\bf q}-$dependent 
susceptibilities $\chi({\bf q}, \omega_{n})$ as the spatial Fourier 
transform of ${\overline{\chi}}_{ij}(\omega_{n})$. 
Introducing $J({\bf q})$, the Fourier transform of the magnetic exchanges 
$J_{ij}^{AF}/\sqrt{z}$, the Bethe-Salpeter Eq.~(\ref{EquationBS}) can be written as 
\begin{equation}
\chi({\bf q}, \omega_{n})
=
\left[
{\Pi}_{0}^{-1}(\omega_{n})
+
J({\bf q})
\right]^{-1}~. 
\label{chideqappendix}
\end{equation}
The AF part ${\cal K}_{AF}$ of the Kernel is expressed in terms of 
the averaged susceptibilities using the definition [cf. Eq.(\ref{finalkernel})]
\begin{eqnarray}
{\cal K}_{AF}(\omega_{n})
&=&
\frac{1}{z}\sum_{ij}
 J_{0i}^{AF}J_{j0}^{AF}
\left[
\overline\chi_{ij}(\omega_{n})
\right. \nonumber \\
&&\left.
-\overline\chi_{i0}(\omega_{n})\overline\chi_{0j}(\omega_{n})/
\overline\chi_{00}(\omega_{n})
\right]~. 
\end{eqnarray}
We rewrite this expression in momentum space: 
\begin{eqnarray}
{\cal K}_{AF}(\omega_{n})
&=&
\sum_{\bf q}J^{2}({\bf q})\chi({\bf q}, \omega_{n})
\nonumber \\
&-&\left[\sum_{\bf q}J({\bf q})\chi({\bf q}, \omega_{n})\right]^{2}/
\chi_{loc}(\omega_{n})~. 
\end{eqnarray}
Using this relation together with Eq.~(\ref{chideqappendix}), and the identity 
$\chi_{loc}(\omega_{n})=\sum_{\bf q}\chi({\bf q}, \omega_{n})$, we find 
\begin{equation}
\Pi_{0}^{-1}(\omega_n)={\cal K}_{AF}(\omega_n)+\frac{1}{\chi_{loc}(\omega_n)}~. 
\end{equation}
An explicit similar 
calculation can be found in Appendix C of Ref.~\onlinecite{smithsi}



\begin{thebibliography}{99}
\bibitem{notespin-liquid}We here adopt the term ``spin-liquid'' to indicate
        a magnetically correlated state, where low dimensionality, spin frustration, 
        and/or quantum fluctuations prevent spin long-range order. In this state
        spin correlations reduce the Curie susceptibility
        to a much weaker power-low (or even logarithmic, see, Ref.
        \onlinecite{sachdevye}) behavior, and the spin fluctuations acquire critical dynamics
        with unusual weight at low frequencies.
\bibitem{nota1suz} Since ${\overline{\chi}}_{ij} 
\sim O(z^{|i-j|/2})$ (as usual $|i-j|$
is the Manhattan distance between sites $i$ and $j$) the second term 
${\overline{\chi}}_{i0}
{\overline{\chi}}_{0j}/
{\overline{\chi}}_{00}\sim O(1/z)$ if both $i$ and $j$ are 
nearest neighbors of site $0$, [as they have to be in Eq. (\ref{completekernel})]. Then,
when the summation can be extended to independent $i$ and $j$ sites,
a $z^2$ factor appears compensating the $1/z$ dependencies. This is not
the case for the second term of the kernel when the disorder interaction
is considered, since in this case there is the additional
constraint $i=j$, and only the 
${\overline{\chi}}_{ii} \sim O(1)$ part of the kernel survives.
\bibitem{notaJAF0} Formally the local model with pure disordered coupling
        is obtained taking the  $J\to 0 $ (implying $\Pi=\chi_{loc}^{-1}$)
        in Eq. (\ref{SGcondition}).
\bibitem{noteharris}
It is worth noticing that this result is reminiscent of the
Harris criterion \cite{harris} establishing the relevance of disorder
on the critical behavior of systems with $D<2/\nu=2(2-\eta)/\gamma$.
Within the EDMFT approach $\eta \equiv 0$ leading to
$D<2/\nu=4/\gamma$. However, the two criteria are physically distinct.
While the Harris criterion arises from arguments based on fluctuations
of the defect distribution on volumes of the order of the correlation
length $\xi$, our argument, which is specific to the EDMFT approach, 
does not apparently
involve any relation between space fluctuations of $J_D$ and $\xi$.
This different physical content is also visible in the fact that 
our criterion does not involve the exponent $\gamma$. Indeed, one
can introduce a ``mass'' term $|T-T_N|^\gamma$ in the denominator 
of the integrand in (\ref{SGcondition}). However, for $D\le 4$
the integral diverges [as $\left(|T-T_N|^\gamma\right)^{\frac{D-4}{2}}$]
for any value of $\gamma$, thereby validating our criterion irrespectively
from $\gamma$.
\bibitem{notemermin}Notice that this is a consequence of the EDMFT
        formalism and, although compatible with, it is not a consequence of
        the Mermin-Wagner theorem. Indeed the same absence of an AF order
        at any finite T would still hold for an Ising model.
\bibitem{note3D} Notice also that at this stage the self-consistent
        expression of the spin self-energy, as well as the related quantities
        $\chi_{loc}$ and ${\cal K}_{AF}$ do not depend on disorder
        and have the same form of a 3D purely AF system. The only role of
        disorder in the paramagnetic phase we consider is to allow the
        application of the SG instability criterium Eqs. (\ref{SGcondition}).
\bibitem{notenomomentum}Notice that in our previous work we also 
        carried out analytic calculations within a large-N expansion.
        Obviously, since the large-N analysis is ``by construction'' unable
        to account for local moment formation, this occurrence was not
        considered in that framework.
\bibitem{notaplot}To allow a comparison of the
curves at different temperatures we plot the curves as functions of
$\tau/\beta$. Due to the bosonic character of the spin excitations,
the curves are obviously periodic with period one in the $\tau/\beta$
variable.
\bibitem{rosch} This peculiar instability has also recently been found within 
        an EDMFT treatment of the t-J model \cite{haule}.
\bibitem{notatstar}Recently we were able to push 
our QMC analysis to lower temperatures $T \lesssim T^*$ in the absence of
disorder. We found that also in this case there are hints of the system
having a tendency to form static local moments. In particular
one can see in Fig. 5 that even the curve for $J_D=0$ displays
a change of slope at low $T$. This is further evidence that the spin
system would like to enter an ordered phase (which is not
allowed by our approach) and shifts spectral weight to larger
frequencies. This shift, however,
is not sufficient to control the exceeding increase of $\chi_{loc}''$ 
at low frequencies and does not prevent the causality catastrophe.
Nevertheless it is useful to notice that any, even weak, additional
physical mechanism leading to a further reduction of $\chi_{loc}''$ 
 might help in avoiding the causality violation. Example in this sense
are given by disorder, geometric frustration, which may prevent
the tendency to magnetic ordering in favor of short-range singlet
correlations.
\bibitem{notaloop}If the $l,m$ sites were common to more than one loop 
decoration, the  
whole path consisting of ``skeleton'' and loops would go along a more constrained
set of sites and would be of higher order in $1/z$. In other words, in the limit of 
infinite coordination it is ``very unlikely'' that loop paths starting from
different sites intersect.
\end{thebibliography}
\end{document}